\documentclass[twocolumn,prl,aps,epsfig,amsmath,amssymb,superscriptaddress,showpacs]{revtex4-2}

\usepackage{graphicx}
\usepackage{dcolumn}
\usepackage{bm}
\usepackage{color}
\usepackage{float}
\usepackage{multirow}
\usepackage{tabularx}
\usepackage{soul}

\begin{document}


\title{Ultrafast measurements under anisotropic strain reveal near equivalence of competing charge orders in TbTe$_3$}

\author{Soyeun Kim}
\affiliation{Stanford PULSE Institute, SLAC National Accelerator Laboratory, Menlo Park, CA 94025, United States}
\affiliation{Stanford Institute for Materials and Energy Sciences, SLAC National Accelerator Laboratory, Menlo Park, CA 94025, USA}

\author{Gal Orenstein}
\affiliation{Stanford PULSE Institute, SLAC National Accelerator Laboratory, Menlo Park, CA 94025, United States}
\affiliation{Stanford Institute for Materials and Energy
Sciences, SLAC National Accelerator Laboratory, Menlo Park, CA 94025, USA}

\author{Anisha G. Singh}
\affiliation{Stanford Institute for Materials and Energy
Sciences, SLAC National Accelerator Laboratory, Menlo Park, CA 94025, USA}

\author{Ian R. Fisher}
\affiliation{Stanford Institute for Materials and Energy
Sciences, SLAC National Accelerator Laboratory, Menlo Park, CA 94025, USA}

\author{David A. Reis}
\affiliation{Stanford PULSE Institute, SLAC National Accelerator Laboratory, Menlo Park, CA 94025, United States}
\affiliation{Stanford Institute for Materials and Energy
Sciences, SLAC National Accelerator Laboratory, Menlo Park, CA 94025, USA}

\author{Mariano Trigo}
\affiliation{Stanford PULSE Institute, SLAC National Accelerator Laboratory, Menlo Park, CA 94025, United States}
\affiliation{Stanford Institute for Materials and Energy
Sciences, SLAC National Accelerator Laboratory, Menlo Park, CA 94025, USA}

\maketitle

{\bf A central problem in condensed matter physics is understanding and controlling emergent phases in complex materials. Unlike hydrostatic pressure and epitaxial strain, uniaxial strain can explicitly affect the lattice symmetry and uncover states with new broken symmetries. This can have a dramatic impact on systems that nearly break rotational symmetries such as nematic electronic order, and competing phases.  Here we combine ultrafast optical measurements with anisotropic strain to study the dynamics of charge density wave (CDW) order in TbTe$_3$. This system hosts two competing orthogonal CDWs in nearly square tellurium planes and can be tuned by uniaxial strain in equilibrium. The measurements show that the a- or c-axis CDW order parameter and its amplitude mode harden with increasing a- or c-axis tensile strain, respectively, indicating that the associated order becomes more stable with increasing tensile strain. This near equivalence of the two orders provides further evidence for an emergent four-fold symmetry of the charge order. More generally, our work demonstrates how the order parameter dynamics under uniaxial strain can uncover the free energy landscape of hidden phases in complex materials.}

\section{introduction}
Interesting phenomena in quantum materials are often found near boundaries between different competing ground states. Understanding the competition between nearly degenerate broken symmetry phases will enable rational control of desirable properties \cite{Basov2017}. Hydrostatic pressure and strain engineering, which can modify exchange couplings and inter-site hopping energies, are particularly fruitful approaches for tuning these competing phases \cite{Ginzburg1963, Degiorgi2007, Zocco2015, Schlom2007, Schlom2014, Xu2020, Ji2019}. 
Compared to hydrostatic pressure or epitaxial strain, uniaxial strain can more effectively lower the lattice symmetry and induce new states with different broken symmetries not accessible by other means. This can have a qualitative impact on systems with competing phases, a prime example being nematic electronic order and its relation to superconductivity \cite{Fradkin2010}.

The rare-earth tri-tellurides, $R$Te$_3$ ($R$ = rare earth ions), provide an attractive platform to realize uniaxial strain-induced phases and control of competing orders. These materials exhibit two competing charge density wave (CDW) instabilities along two perpendicular crystallographic axes with almost identical lattice parameters ($a$ = 4.3081(10) and $c$ = 4.3136(10) for TbTe$_3$ at room temperature \cite{Ruthesis}). Application of isotropic pressure in these materials, such as chemical pressure \cite{Ru2008prb, Hu2014} and hydrostatic pressure \cite{Degiorgi2007, Zocco2015, Kopaczek2022}, have been shown to tune the CDW instability and even enhance superconductivity \cite{Hamlin2009}, suggesting that the equilibrium phase can be tuned with strain. Contrary to the isotropic pressure, anisotropic strain in the $a$-$c$ plane offers a more direct and effective means of tuning the near-degeneracy between the two perpendicular CDWs, because the CDWs are hosted in the near-square, quasi-two-dimensional Te layers \cite{Ruthesis, Anisha2023, Straquadine2022}.

In this work, we present ultrafast reflectivity measurements of the dynamics of the CDW order parameter in TbTe$_3$ under anisotropic strain at room temperature. 
Unlike equilibrium measurements, ultrafast spectroscopy can provide crucial information on the single-particle and collective modes of the CDW \cite{Yusupov2008, Schaefer2014, Demsar1999}. By virtue of its nonequilibrium nature, pump-probe spectroscopy can unveil valuable information inaccessible by other means. For example, nonlinear dynamics of the order parameter under high excitation reveal the anharmonic shape of the potential energy \cite{Yusupov2010, Huber2014, Trigo2021, Orenstein2023}, and provides deeper insights into the broken symmetry phase. 

Recent experiments give support to the near equivalence of the two CDWs in $R$Te$_3$. 
Ultrafast electron diffraction (UED) studies have shown an emergent CDW with wavevector in the $a$ direction after photoexcitation quenches the stable CDW with wavevector along the $c$-axis \cite{Kogar2019}. 
Equilibrium x-ray diffraction measurements showed a similar rotation of the CDW wavevector by 90 degrees (from the $c$-axis to the $a$-axis) with tensile strain along the $a$-axis \cite{Straquadine2022, Anisha2023, Gallo2023}. 
These results suggest an intimate relationship between the $a$- and $c$-axes instabilities, near equivalence between these two directions, and hint of an emergent symmetry proposed in ErTe$_3$ \cite{Anisha2023}.

\begin{figure*}[ht]
\begin{center}
\includegraphics[width = 159 mm]{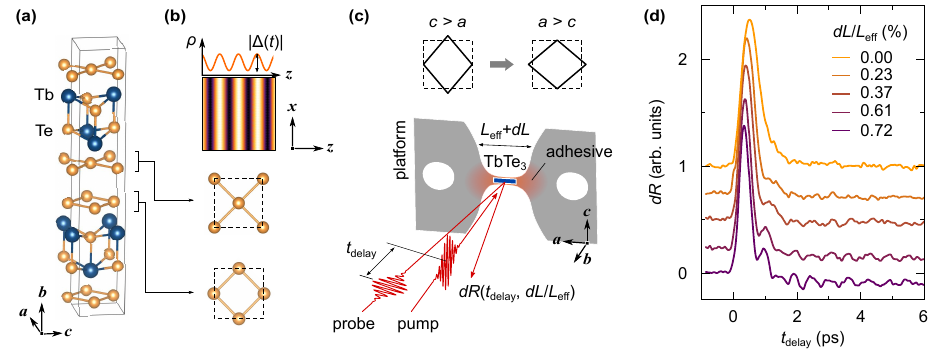}
\caption{\textbf{(a)} The crystal structure of TbTe$_3$ in the $Cmcm$ space group, featuring nearly tetragonal Te bilayers that separate TbTe blocks. \textbf{(b)} A schematic representation of the charge density ($\rho$) amplitude ($|\Delta|$) in real space. The charge-density-wave (CDW) phase forms along the $c$-axis at room temperature, with a cross-section along the $c$-axis displayed at the top of the image. \textbf{(c)} An illustration of the strain-dependent transient reflectivity setup. The sample is attached to the bowtie platform and experiences the anisotropic strain along the $a$-axis. The crystal axes are shown with arrows. $L_\text{eff}$ denotes the effective distance, and $dL$ the applied displacement. Two near-normal 800 nm pump and probe beams are incident onto the sample. \textbf{(d)} The transient time traces for various $dL$ values, each displayed with an offset.}
\label{fig-1}
\end{center}
\end{figure*}

Our ultrafast reflectivity measurements on TbTe$_3$ show a stiffening of the CDW amplitude mode (AM) for a strong enough tensile strain along the $a$-axis, which is known to reorient the CDW wavevector by 90 degrees from the $c$- to the $a$-axis \cite{Anisha2023, Gallo2023}. This stiffening indicates that strain further stabilizes the order parameter of the rotated CDW for $a > c$. 
On the other hand, tensile strain along the $c$-axis ($c > a$), also stiffens the original AM. 
These observations indicate a strong similarity between the potential energy surface of the two CDWs and further support the proposed near equivalence between the $c$ and $a$ orders. 

To further interpret the results, we analyze the reflectivity dynamics within the framework of the time-dependent Ginzburg-Landau model \cite{Yusupov2010, Schaefer2014, Trigo2019}.
Our analysis of the coherent order parameter dynamics indicates that the system stiffens and develops a deeper free energy minimum for either $a/c < 1$ or $a/c > 1$, which further stabilizes the $c$ and $a$ order, respectively. The near-identical behavior of the two strain regimes is a signature of the near equivalence of the two CDWs.

TbTe$_3$ crystallizes in the $Cmcm$ space group (No. 63) and undergoes a CDW phase transition at $T_\text{CDW}$ = 336 K \cite{Ru2008prb}. As in other series of the $R$Te$_3$ family, the system is orthorhombic as depicted in Fig. \ref{fig-1}(a). At room temperature, the $c$ lattice constant is larger than $a$ by 0.13\% and the CDW wavevector $\bm{q}_\text{CDW}$ is oriented along the $c$-axis, as in Fig. \ref{fig-1}(b). The difference between $a$ and $c$ reduces at higher temperatures and the CDW becomes unstable.

The small in-plane anisotropy is considered crucial for determining the direction of the CDW in $R$Te$_3$.
Recent x-ray diffraction studies on the anisotropic-strain dependence on ErTe$_3$ and TbTe$_3$ have demonstrated the CDW can be reoriented by controlling the lattice constant ratio $a/c$ \cite{Anisha2023, Gallo2023}. Because applying strain does not remove the glide plane, it does not break the $C_4$ rotation symmetry. Nevertheless, x-ray diffraction measurements showed that the CDW wavevector rotates by 90$^\circ$ at sufficiently large $a/c$ \cite{Anisha2023}.
Resistivity measurements under strain confirmed that the transition temperature increases as $a/c$ deviates further from 1 for both $a/c < 1$ and $a/c > 1$ \cite{Anisha2023, Gallo2023}, suggesting that strain stabilizes the respective $c$ or $a$ CDW order. 

\section{results}
\textbf{Anisotropic strain control in TbTe$_3$.} To manipulate the CDW state, an anisotropic strain field was applied along the $a$-axis in TbTe$_3$. The bulk TbTe$_3$ sample was attached to the neck of the bowtie-shaped platform as illustrated in Fig. \ref{fig-1}(c) following the method in \cite{Anisha2023}. Each end of the platform was fixed to a commercial strain cell device that provided tensile strain in the sample, which was monitored with the displacement $dL$ of the cell \cite{Razorbill}. The applied strain is estimated using the ratio between $dL$ and the effective length of the platform $L_\text{eff}$ (For more details see the Methods).

\textbf{Transient reflectivity measurement.}
Figure \ref{fig-1}(d) presents the strain-dependent time traces of the transient reflectivity at room temperature. The scans were initiated from the maximum $dL / L_\text{eff}$ and sequentially lowered (bottom to top traces) to minimize the extrinsic change in the reflectivity, such as crack formation.
The general features observed in the traces depicted in Fig. \ref{fig-1}(d) align with findings from previous transient reflectivity studies on $R$Te$_3$ \cite{Yusupov2008, Yusupov2010, Kusar2011, Zong2019prl, Trigo2019}. The shape of the initial peak relates to the initial order parameter dynamics, and the rise time of the peak was interpreted as the time needed to fully suppress the CDW order \cite{Zong2019prl}. Phenomenological Ginzburg-Landau models based on quartic potentials \cite{Sun2020} successfully describe the trend in transient reflectivity at different temperatures and over a wide range of pump fluences, including at high fluence where nonlinearities in the lattice motion become important \cite{Yusupov2010, Trigo2019, Trigo2021, Maklar2021}.

\begin{figure}[t!]
\begin{center}
\includegraphics[width = 79 mm]{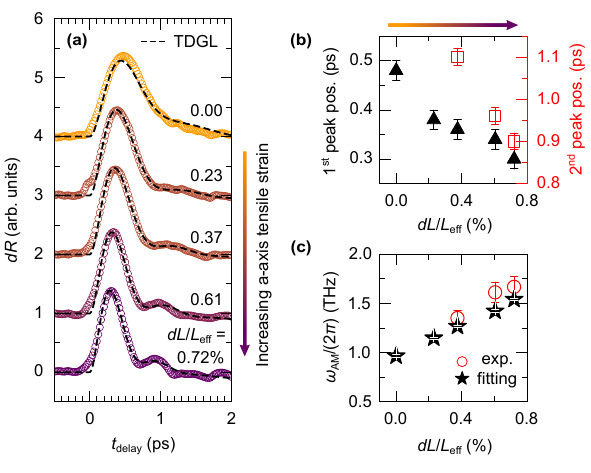}
\caption{\textbf{(a)} Transient reflectivity traces under applied $a$-axis tensile strain. Traces are offset vertically for clarity. The incident pump fluence was kept at 9 $\mu$J/cm$^2$ for all traces. The fit using the time-dependent Ginzburg-Landau (TDGL) model with constrained $\eta_F$ (= 0.1) is overlaid with dashed lines. \textbf{(b)} The temporal positions of the first and second peaks for traces in (a). \textbf{(c)} The $\omega_\text{AM}$ frequency estimated from the time separation of the first two peaks are shown with the open symbol (labeled ``exp."), and the fit results from the TDGL model are represented with star symbols (labeled ``fitting").}
\label{fig-2}
\end{center}
\end{figure}

We now investigate how anisotropic strain affects the order parameter dynamics. When no strain is applied (top trace in Fig. \ref{fig-1}(d); $dL / L_\text{eff}$ = 0\%), the time trace shows a smooth peak of about 0.5 ps after the pump. A second hump starts to appear at 1.2 ps and becomes sharper with increasing $dL / L_\text{eff}$ (lower traces in Fig. \ref{fig-1}(d)). Additional oscillating features become discernible for the time range of about 1.5 to 6 ps, which are prominent at $dL / L_\text{eff}=0.72\%$.

The early dynamics of the reflectivity, $i.e.,$ the initial peak and the subsequent second hump, are primarily influenced by the CDW AM \cite{Trigo2019} and dynamic critical slowing down \cite{ Zong2019prl}. In the signal after $1.5$ ps, coherent oscillations with a frequency of 1.65 THz from an optical phonon coupled to the AM~\cite{Lavagnini2008, Yusupov2008} become predominant. In our analysis below, we focus on the behavior at $t < 1.5$~ps, which reflects the order parameter dynamics in the dynamic critical slowing down regime.

The initial peak shows dramatic change with $dL / L_\text{eff}$, as shown in Fig. \ref{fig-2}(a). With increasing tensile strain along the $a$-axis (increasing $dL / L_\text{eff}$), the peak position ($i.e.,$ the rise time) shifts to earlier times (Fig. \ref{fig-2}(b)). The peak intensity first increases and then decreases slightly, and the peak shape sharpens. 
Importantly, the oscillations of the order parameter become more pronounced at higher $dL / L_\text{eff}$ (see second cycle at $t_\text{delay}=1$~ps in $dL/L_\mathrm{eff} = 0.72$\%). This occurs because of an increase in the frequency of the AM at higher strain.

The rise time of the initial peak $\tau_\text{rise}$ is related to the period of the AM, which diverges (the AM frequency vanishes) at the critical point in equilibrium
\cite{Zong2019prl}. Thus, the decrease in $\tau_\text{rise}$ with increasing $dL / L_\text{eff}$ observed here implies that the order parameter becomes more robust with increasing $dL / L_\text{eff}$ and that the system is pushed away from the critical point by further stabilizing the ordered phase. The initial peak dynamics show a similar CDW stabilizing trend when tensile strain is applied along the $c$-axis (see Fig. S3 in Supplementary Information \cite{SI}). 
Importantly, although their behavior is similar, these two situations correspond to distinct orthogonal CDW orders ($a$ and $c$, respectively)\cite{Anisha2023}.

\begin{figure}[t!]
\begin{center}
\includegraphics[width = 79 mm]{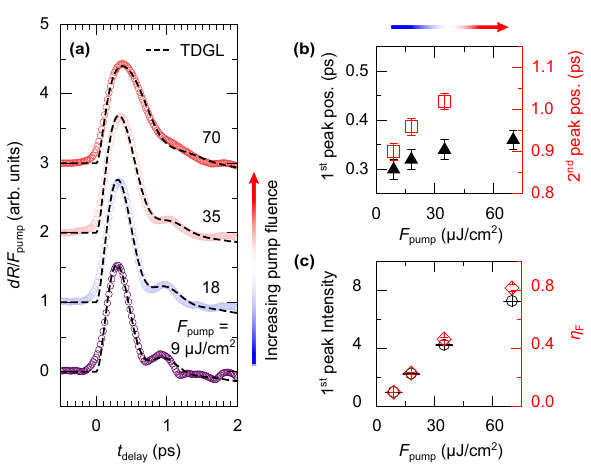}
\caption{\textbf{(a)} Transient reflectivity time traces under applied $a$-axis tensile strain ($dL/L_\text{eff}$ = 0.72\%). Each trace is normalized by fluence and offsets vertically. The fit using the TDGL model with constrained $\omega_\text{AM}/(2\pi)$ (= 1.54 THz) and $A$ (= 0.0015) is overlaid with dashed lines. \textbf{(b)} The positions of the first and second peaks in the traces. \textbf{(c)} The maximum amplitude of $dR$ for each fluence is shown with circle symbols. The $\eta_F$ values obtained from the TDGL model fit are represented with diamond symbols.}
\label{fig-3}
\end{center}
\end{figure}

\begin{figure*}[t!]
\begin{center}
\includegraphics[width = 140 mm]{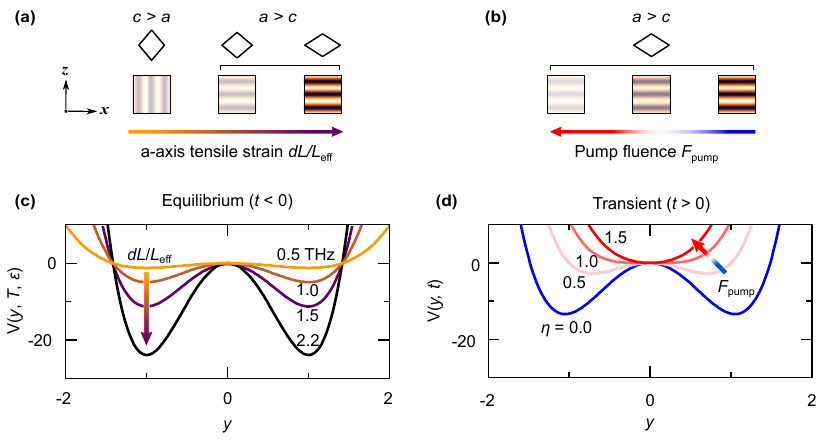}
\caption{\textbf{(a-b)} Schematic diagrams of the CDW affected by \textbf{(a)} increasing the $a$-axis tensile strain and \textbf{(b)} increasing pump fluence. The arrows in \textbf{(a,b)} indicate the increasing directions of $dL/L_\text{eff}$ or $F_\text{pump}$. \textbf{(c)} The equilibrium Ginzburg-Landau potential $V(y, T, \varepsilon)$ shape for the representative $\omega_\text{AM}/(2\pi)$ values. \textbf{(d)} The transient Ginzburg-Landau potential $V(y, t)$ of $\omega_\text{AM}/(2\pi)$ = 1.5 THz for the representative $\eta_F$ values.}
\label{fig-4}
\end{center}
\end{figure*}

\textbf{Time-dependent Ginzburg-Landau model.}
To gain further insight into the strain-dependency in the transient reflectivity, we analyze the traces using the time-dependent Ginzburg-Landau (TDGL) model for second-order phase transitions, which provides a good phenomenological description of the CDW order parameter dynamics in $R$Te$_3$ \cite{Yusupov2010, Trigo2019, Zong2019prl, Trigo2021, Maklar2021}. 
Following Ref. \cite{Trigo2019}, the model is constructed by assuming a real order parameter $y$, which is the amplitude of the CDW distortion at $\bm{q}_\text{CDW}$ relative to its equilibrium value at a given temperature. Note that due to the normalization of the order parameter, the minimum of $V(y)$ is at $y=\pm 1$ in equilibrium for all temperatures, while with the parametrization below, the fluence can change the position of the minimum away from $y=\pm 1$. The effective potential in terms of $y$ is
\begin{equation}
     V(y) = \frac{\omega_\text{AM}^2}{8} \Big( 2(\eta-1)y^2 + y^4 \Big),
\end{equation}\label{eq-V} and the dynamics of $y$ are governed by the equation of motion \begin{equation}
  \frac{2}{ \omega_\text{AM}^2 } \ddot{y} + \big(\eta-1 \big)y + y^3 + \frac{2 \gamma}{\omega_\text{AM}^2}\dot{y} = 0.
\end{equation}\label{eq-EOM} 

Here $\omega_\text{AM}$ is the AM angular frequency, $\eta$ is a parameter that controls the stability of the CDW which depends on the pump fluence and time \cite{Trigo2019}, and $\gamma$ is phenomenological damping. Note that $\omega_\text{AM}$ is a function of temperature and applied strain. This way, changes in the potential energy with strain can be tracked by monitoring changes in the frequency $\omega_\text{AM}$, or equivalently, $\tau_{\rm rise}$. Here $\eta = 0$ corresponds to the CDW ground state in equilibrium ($t<0$) and $\eta > 0$ describes the free energy of the transient high-symmetry \cite{Trigo2019, Yusupov2010} and metastable phases \cite{Maklar2021}. To introduce the effect of the optical pump, we take $\eta(t) = \eta_F (e^{-t/\tau}-\eta_m)/(1-\eta_m) \Theta(t)$, where $\Theta(t)$ is a step function, $\tau$ is the decay time of the electronic excitation, $\eta_F$ parameterizes the pump fluence, and the offset $\eta_m$ captures the long term change in the excitation before the system fully equilibrates. Further detailed derivation of the model and relevant parameters are provided in the Supplementary Information \cite{SI}.

The order parameter $y(t)$ dynamics are related to the transient reflectivity and directly compared to the experiment. The reflectivity is insensitive to the sign (phase) of the order parameter, and thus by symmetry, the expansion in terms of in $y$ is quadratic to the lowest order \cite{Ginzburg1963, Trigo2019}. Therefore we fit the time traces to $A(1-y^2(t)$), $A$ being a signal amplitude.

\section{discussion}
Here we compare the order parameter dynamics $y(t)$ in the TDGL model to the experimental transient reflectivity. Figure \ref{fig-2} presents the fit results of the TDGL model to reflectivity traces for the AM dynamics at $t_\text{delay} < 1.5$ ~ps. The fitted curves shown in dashed lines in panel (a) capture the main features of the experimental traces, including the trend with increasing strain. The fitted $\omega_\text{AM}$ values in (c) show a gradual increase with increasing $dL / L_\text{eff}$. Because the pump fluence was fixed to 9 $\mu$J/cm$^2$ for measurements in Fig. \ref{fig-2}, the $\eta_F$ was fixed. Note the other parameters are comparable for different $dL / L_\text{eff}$ data, with $\gamma \sim$ 6 THz and $\tau \sim$ 0.5 ps (The fit parameters are provided in Table S1 in the Supplementary Information \cite{SI}). 

The frequency $\omega_\text{AM}/(2\pi)$ can be directly compared with the inverse of the single-cycle period between two peak positions. Such values for high $dL / L_\text{eff}$ traces that have a definite peak position in the second oscillation are overlaid (circle symbols) in Fig. \ref{fig-2}(c), which matches well with the TDGL fit results (start symbols). These observations, the increase in the $\omega_\text{AM}$ with $dL / L_\text{eff}$, signifies that the $a$-axis strain pushes the potential energy into deeper double-well form as shown schematically in Fig. \ref{fig-4}(a), hence stabilizing the CDW, which is rotated 90$^\circ$ to that of the unstrained CDW \cite{Anisha2023}.

In Fig. \ref{fig-3}, we verified the pump fluence ($F_\text{pump}$) dependence of the strain-stabilized CDW at $dL / L_\text{eff}$ = 0.72\%. As $F_\text{pump}$ increases, the AM oscillation becomes suppressed, and the rise time increases (triangle symbols in panel (b)). The maximum amplitude of $dR$ increases linearly with $F_\text{pump}$ as circle symbols in panel (c), showing a similar trend in unstrain LaTe$_3$ \cite{Zong2019prl} and SmTe$_3$ \cite{Trigo2019} at fluences below the dynamical slowing down regime. Note that for the unstrained system, it is expected that increasing the pump fluence ($F_\text{pump}$) destabilizes the CDW order \cite{Zong2019prl, Maklar2021, SI}. 

The TDGL model fit for the pump fluence dependence is overlaid in Fig. \ref{fig-3}(a) with dashed lines. Since the applied strain and temperature are constant here, we constrained $\omega_\text{AM}$ and $A$ to the fitted values for the lowest fluence trace (9 $\mu$J/cm$^2$) in Fig. \ref{fig-2}. Both the fit parameters $\eta_F$ (diamond symbols in panel (c)) and the amplitude $A$ (now shown) showed an increasing trend, consistent with the initial peak intensity growth with $F_\text{pump}$. The values of $\tau$ and $\gamma$ also increase for higher $F_\text{pump}$, while there is almost no change in $\eta_m$ (The fit parameters are provided in Table S2 in the Supplementary Information \cite{SI}). In general, the $F_\text{pump}$ dependence for the state with the largest strain is consistent with the known behavior in unstrained crystals \cite{Zong2019prl, Trigo2019}. 

Figure \ref{fig-4} summarizes the behavior of the Ginzburg-Landau potential with the $a$-axis tensile strain and pump fluence, extracted from the fits in Figs. \ref{fig-2} and \ref{fig-3}, respectively. As mentioned earlier, prior strain studies on ErTe$_3$ and TbTe$_3$ demonstrated that the CDW wavevector is rotated from the $c$- to the $a$-axis when $a>c$ \cite{Anisha2023, Gallo2023} (panel (a)). As the in-plane anisotropy grows, the reoriented CDW is further stabilized which makes the potential well deeper (panel (c)), as evidenced by the observed decrease in $\tau_\text{rise}$ and increase in the fitted $\omega_\text{AM}$ with increasing $dL / L_\text{eff}$ (Figs. 2(b,c)). 
On the other hand, the strain-induced reoriented CDW is destabilized (panel (b)) as the incident pump transiently drives the potential to become shallower (panel (d)), similar to the unstrained CDW. The fact that both CDWs can be modeled with similar parameters within TDGL suggests that the original (unstrained) and reoriented CDW are virtually indistinguishable and both have comparable free energies. This suggests they are manifestations of the same free energy instability and further supports the interpretation that the original (with wavevector along $c$) and reoriented (with wavevector along $a$) CDW orders are nearly equivalent \cite{Anisha2023}.

To conclude, we used ultrafast optical spectroscopy to probe the dynamics of the order parameter in TbTe$_3$ under anisotropic strain. The response when $a/c < 1$ and $a/c > 1$ is nearly identical and both CDWs can be modeled with the same Ginzburg-Landau potential despite the intrinsically orthorhombic nature of the underlying lattice, which further supports the near equivalence of the two CDWs.
These results demonstrate a versatile tool combining ultrafast spectroscopy with a uniaxial strain tuning device to probe the potential energy surfaces and low energy excitations of competing orders in quantum materials.

\section{Method}
\textbf{Sample preparation}
Single crystals of TbTe$_3$ were synthesized via a self-flux technique as described in Ref. \cite{Ruthesis}. An anisotropic strain field was applied along the crystal $a$-axis using a CS-130 Razorbill strain cell device \cite{Razorbill}. As depicted in Fig. \ref{fig-1}, the bulk TbTe$_3$ sample was attached to the neck of the bowtie-shaped platform, following the approach in the ErTe$_3$ study \cite{Anisha2023}. The sample was cleaved to a thickness of 5-10 $\mu$m to expose a fresh surface and installed onto the strain cell device for the reflectivity measurements. The effective length $L_\text{eff}$ of the bowtie neck where the strain is active is 3.47 mm \cite{Anisha2023, Park2020}, and the displacement change $dL$ in the strain cell device is obtained from the internal capacitor equipped inside the cell \cite{Razorbill}. It is noteworthy that for most values of $dL$ we used, the platform approaches or exceeds the plastic deformation limit of the titanium bowtie. This indicates the actual strain applied on the sample that affects the $a/c$ may not have a linear relation with $dL$. Further details are provided in Ref. \cite{Anisha2023} and in the Supplementary Material \cite{SI}.

\textbf{Ultrafast optical measurements}
The transient reflectivity time traces were obtained from an optical pump-probe setup using a Coherent RegA laser with 800 nm wavelength at a 250 kHz repetition rate. The probe (pump) beam had a $1/e^2$ width of 40 (170) $\mu$m with a near-normal incident angle, which passed through a neutral density filter to change the fluence. The polarization of the probe and pump beams were orthogonal to each other. The photon energy 1.55 eV is much higher than the CDW optical gap of TbTe$_3$ at room temperature (220 meV) \cite{Hu2014}. The pulse duration of the pump and probe beams was around 70 fs at the sample position, and the pump-probe delay was controlled with a mechanical delay stage.

\vspace{4mm}
\noindent\textbf{Acknowledgments}\\
S.K., G.O., A.G.S., I.R.F., D.A.R., and M.T. were supported by the US Department of Energy, Office of Science, Office of Basic Energy Sciences through the Division of Materials Sciences and Engineering under Contract No. DE-AC02-76SF00515. G.O. acknowledges support from the Koret Foundation. 

\vspace{4mm}
\noindent\textbf{Author contributions}\\ S.K., G.O., and M.T. designed the experiment. S.K. and G.O. constructed the setup and performed experiments. S.K. analyzed the results. A.G.S. synthesized and characterized the sample. I.R.F., D.A.R., and M.T. gave insights to the understanding and analysis of the data. S.K. and M.T. wrote the manuscript with input from all authors.

\vspace{4mm}
\noindent\textbf{Competing interests}\\ The authors declare no competing interests.

\vspace{4mm}

\noindent\textbf{Data availability}\\ All relevant data are available upon request.

\bibliographystyle{apsrev4-2}
\bibliography{References.bib}

\end{document}